\documentclass[twocolumn,preprintnumbers,amsmath,amssymb]{revtex4}
\usepackage{graphicx}
\usepackage{dcolumn}
\usepackage{bm}

\begin{document}
\title{\bf Vehicular traffic flow at a non-signalised intersection }
\author{M. Ebrahim Foulaadvand ~\footnote{Corresponding author: foolad@iasbs.ac.ir}}
\affiliation{Department of Nano-Science,
 Institute for Studies in Theoretical Physics and Mathematics (IPM),
P.O. Box 19395- 5531,Tehran, Iran and Department of Physics, Zanjan
University, P.O. Box 45195-313, Zanjan,  Iran.}
\author{Sommayeh Belbaasi}
\affiliation{Department of Physics, Zanjan University, P.O. Box
45195-313, Zanjan,  Iran.}
\date{\today}
\begin{abstract}

We have developed a modified Nagel-Schreckenberg cellular automata
model for describing a conflicting vehicular traffic flow at the
intersection of two streets. No traffic lights control the traffic
flow. The approaching cars to the intersection yield to each other
to avoid collision. Closed boundary condition is applied to the
streets. Extensive Monte Carlo simulations is taken into account to
find the model characteristics. In particular, we obtain the
fundamental diagrams and show that the effect of interaction of two
streets can be regarded as a dynamic impurity located at the
intersection point. Our results suggest that yielding mechanism
gives rise to a high total flow throughout the intersection
especially in the low density regime. In some ranges of densities,
yielding mechanism even improves and regulates the flow in
comparison to the absence of perpendicular flow.\\

\end{abstract}

\maketitle
\section{{Introduction}}
Modeling the dynamics of vehicular traffic flow has constituted the
subject of intensive research by statistical physics and applied
mathematics communities during the past years
\cite{kernerbook,schadrev,helbingrev,klar,bellom}. In particular,
cellular automata approach have provided the possibility to study
various aspects of these truly non-equilibrium systems which still
are of current interest \cite{tgf01,tgf03,tgf05}. Besides various
theoretical efforts aiming to understand the basic principles
governing the spatial-temporal structure of traffic flow,
considerable attempts have been made towards realistic problems
involving optimization of vehicular traffic flow. While the existing
results in the context of highway traffic seem to need further
manipulations in order to find direct applications, researches on
{\it city traffic} have more feasibility in practical applications
\cite{bml,nagatani,tadaki1,tadaki2,cuesta,torok,freund,cs,brockfeld}.
We believe that optimisation of traffic flow at a single
intersection is a substantial ingredient for the task of global
optimisation of city networks \cite{chitur}. Isolated intersections
are fundamental operating units of complex city networks and their
thorough analysis would be inevitably advantageous not only for
optimisation of city networks but also for local optimization
purposes. Recently, physicists have paid notable attention to
controlling traffic flow at intersections and other traffic
designations such as roundabouts
\cite{foolad1,foolad2,foolad3,foolad4,helbing1,helbing2,helbing3,ray,xiong,gershenson}.
In this respect, our objective in this paper is to study another
aspect of conflicting traffic flow at intersections. In principle,
the vehicular flow at the intersection of two roads can be
controlled via two distinctive schemes. In the first scheme, which
is appropriate when the density of cars in both roads are low, the
traffic is controlled without traffic lights. In the second scheme,
signalized traffic lights control the flow. In the former scheme,
approaching car to the intersection yield to traffic at the
perpendicular direction by adjusting its velocity to a safe value to
avoid collision. According to driving rules, the priority is given
to the nearest car to the intersection. It is evident that this
scheme is efficient if the density of cars is low. When the density
of cars increases, this method fails to optimally control the
traffic and long queues may form which gives rise to long delays. At
this stage the implementation of the second scheme i.e.; utilizing
traffic lights is unavoidable. Therefore it is a natural and
important question to find out under what circumstances the
intersection should be controlled by traffic lights? More concisely,
what is the critical density beyond which the non-signalised schemes
begins to fail. In order to capture the basic features of this
problem, we have constructed a cellular automata model describing
the above dynamics. This paper has the following layout. In section
II, the model is introduced and driving rules are explained. In
section III, the results of the Monte Carlo simulations are
exhibited. Concluding remarks and discussions ends the paper in
section IV.

\section{ Description of the Problem }

We now present our CA model. Consider two perpendicular one
dimensional closed chains each having $L$ sites. The chains
represent urban roads accommodating unidirectional vehicular
traffic flow. They cross each other at the sites
$i_1=i_2=\frac{L}{2}$ on the first and the second chain
respectively. With no loss of generality we take the direction of
traffic flow in the first chain from south to north and in the
second chain from east to west (see Fig.1 for illustration). The
discretisation of space is such that each car occupies an integer
number of cells denoted by $L_{car}$. The car position is denoted
by the location of its head cell. Time elapses in discrete steps
of $\Delta t$ $sec$ and velocities take discrete values
$0,1,2,\cdots, v_{max}$ in which $v_{max}$ is the maximum velocity
of cars.

\begin{figure}
\centering
\includegraphics[width=7.5cm]{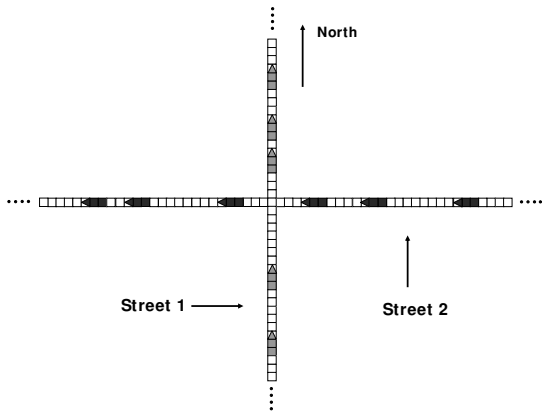}
\caption{ Intersection of two uni-directional streets. } \label{fig:bz2}
\end{figure}

To be more specific, at each step of time, the system is
characterized by the position and velocity configurations of
cars. The system evolves under the Nagel-Schreckenberg (NS)
dynamics \cite{ns}. Let us briefly explain the NS updating rules
which synchronously evolve the system state from time $t$ to
$t+1$. We denote position, velocity and space gap of a typical
car at timestep t by $x^{(t)},v^{(t)}$ and $g^{(t)}$
respectively. The same quantities for its leading car are
correspondingly denoted by $x_l^{(t)},v_l^{(t)}$ and $g_l^{(t)}$.
We recall that gap is defined as the distance between front
bumper of the follower to the rear bumper of its leading. More
precisely, $g(t)=x_l(t)-x(t)-L_{car}$. Concerning the above
considerations, the following updating sub steps evolve the
position

and the velocity of each car in parallel.\\

1) Acceleration:\\

$v^{(t+1/3)}:= min(v^{(t)}+1, v_{max})$\\

2) Velocity adjustment :\\

$v^{(t+2/3)}:=min(g^{(t+1/3)}, v^{(t+1/3)})$\\

3) Random breaking with probability $p$:\\

if random $< p$ then $v^{(t+1)}:=max(v^{(t+2/3)}-1,0)$\\

4) Movement : $x^{(t+1)} :=x^{(t)}+ v^{(t+1)}$ \\

The yielding dynamics in the vicinity of the intersection is
implemented by introducing a safety distance $D_s$. The
approaching cars (nearest cars to the crossing point
$i=\frac{L}{2}$) should yield to each other if their distances to
the crossing point, denoted by $d_1$ and $d_2$ for the first and
second street respectively, are both less than the safety distance
$D_s$. In this case, the movement priority is given to the car
which is closer to the crossing point. This car adjust its
velocity as usual with its leading car. On the contrary, the
further car, which is the one that should yield, brakes
irrespective of its direct gap. The simplest way to take into
this cautionary braking is to adjust the gap with the crossing
point itself. This implies that the yielding car sees the
crossing point as a hindrance. In this way, the model is
collision-free. Figure two illustrates the situation.

\begin{figure}
\centering
\includegraphics[width=7.5cm]{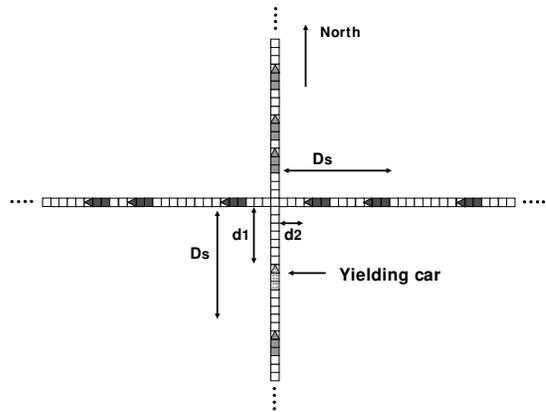}
\caption{ Two approaching cars to the intersection yield to each
other provided their distances to the crossing point are both less
than a safety distance. } \label{fig:bz2}
\end{figure}

Let us now specify the physical values of our time and space
units. Ignoring the possibility of existence of long vehicles
such as buses, trucks etc, the length of each cell is taken to be
as 4.5 metres which is the typical bumper-to-bumper distance of
cars in a waiting queue. Therefore the spatial grid $\Delta x$
equals to $\frac{4.5}{L_{car}}~m$. We take the time step $\Delta
t=1~s$. Furthermore, we adopt a speed-limit of $75~ km/h$. The
value of $v_{max}$ is so chosen to give the speed-limit value
$75~km/h$ or equivalently $21~m/s$ . In this regard, $v_{max}$ is
given by the integer party of $21 \times L_{car}/4.5$. For
instance, for $L_{car}=5$, $v_{max}$ equals $23$. In addition,
each discrete increments of velocity signifies a value of $\Delta
v=\frac{4.5}{L_{car}} m/s $ which is also equivalent to the
acceleration. For $L_{car}=5$ this gives the comfort acceleration
$0.9 ~\frac{m}{s^2}$. Moreover, we take the value of random
breaking parameter at $p=0.1$. In the next section, the
simulation results of the above-described dynamics is presented.

\section{ Monte Carlo simulation}

The streets sizes are equally taken as $L_1=L_2=1350~m$ and the
system is update for $10^6$ time steps. After transients, two
streets maintain steady-state currents, defined as the number of vehicles
passing from a fixed location per a definite time interval, denoted by $J_1$ and $J_2$
. They are functions of the global densities $\rho_1=\frac{N_1
\times L_{car}}{L_1}$ and $\rho_2=\frac{N_2 \times L_{car}}{L_2}$
where $N_1$ and $N_2$ are the number of vehicles in the first and
the second street respectively. We kept the global density at a
fixed value $\rho_2$ in the second street and varied $\rho_1$.
Figure (3) exhibits the fundamental diagram of the first street
i.e.; $J_1$ versus $\rho_1$.

\begin{figure}
\centering
\includegraphics[width=7cm]{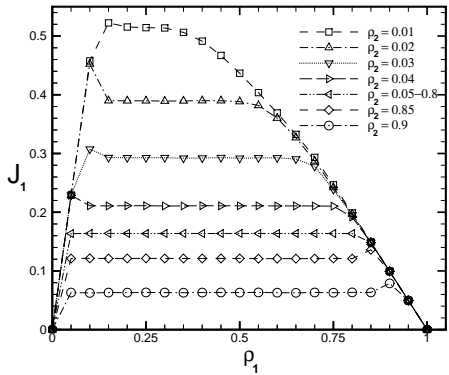}
\caption{$J_1$ vs $\rho_1$ for various values of $\rho_2$.
$D_s=25~m$ and $L_{car}=5$ cells. The road length is $1350$ metre. }
\label{fig:bz2}
\end{figure}

It is observed that for small densities $\rho_2$ up to $0.05$,
$J_1$ rises to its maximum value, then it undergoes a short rapid
decrease after which a lengthy plateau region, where the current
is independent of $\rho_1$, is formed. Intersection of two chains
makes the intersection point appear as a site-wise dynamical
defective site. It is a well-known fact that a local defect can
affect the low dimensional non-equilibrium systems on a global
scale
\cite{lebowitz,evans1,barma1,kolomeisky1,krug,chou,lakatos1,foolad5,kerner}.
This has been confirmed not only for simple exclusion process but
also for cellular automata models describing vehicular traffic
flow \cite{chung,yukawa}. Analogous to static defects, in our
case of dynamical impurity, we observe that the effect of the
site-wise dynamic defect is to form a plateau region $ \rho \in
[\rho_{-}, \rho_{+}]$ in which $\Delta=\rho_{+}-\rho_{-}$ is the
extension of the plateau region in the fundamental diagram. The
larger the density in the perpendicular chain is, the more strong
is the dynamic defect. For higher $\rho_2$, the plateau region is
wider and the current value is more reduced.
After the plateau, $J_1$ exhibits linear decrease versus
$\rho_1$ in the same manner as in the fundamental diagram of a
single road. In this region which corresponds to $\rho_1 >
\rho_{+}$ the intersecting road imposes no particular effect on
the first road. Increasing $\rho_2$ beyond $0.05$ gives rise to
substantial changes in the fundamental diagram. Contrary to the
case $\rho_2 <0.05$, the abrupt drop of current after reaching
its maximum disappears for $\rho_2>0.05$ and $J_1$ reaches its
plateau value without showing any decrease. The length and height
of the plateau does not show significant dependence for $\rho_2
\in [0.05,0.8]$. This marks the efficiency of the non signalised
controlling mechanism in which the current of each street is highly
robust over the density variation in the perpendicular street.
When $\rho_2$ exceeds $0.8$, the plateau undergoes changes. Its
length increases whereas its height decreases.
We now consider the flow characteristics in the second street.
Although the global density is constant in street $2$ its
current $J_2$ is affected by density variations in the first
street. In Figure (4) we depict the behaviour of $J_2$ versus $\rho_1$.

\begin{figure}
\centering
\includegraphics[width=7cm]{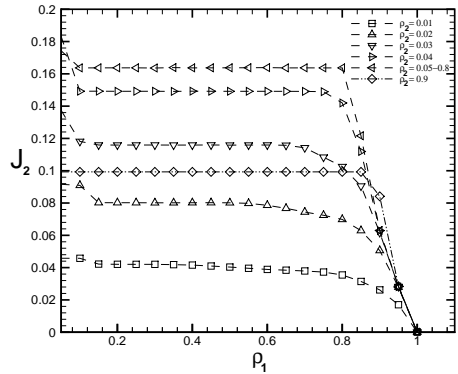}
\caption{ $J_2$ vs $\rho_1$ for various values of $\rho_2$.
$D{s}=25~ m$ and $L_{car}=5$ cells. The road length is $1350$ metre.
} \label{fig:bz2}
\end{figure}

For each value of $\rho_2$, the current $J_2$ as a
function of $\rho_1$ exhibits three regimes. In the first regime
in which $\rho_1$ is small, $J_2$ is a decreasing function of
$\rho_1$. Afterwards, $J_2$ reaches a plateau region (second
regime) which is approximately extended over the region $\rho_1
\in [0.1,0.8]$. Eventually in the third regime, $J_2$ exhibits
decreasing behaviour towards zero. Analogous to $J_1$, the
existence of wide plateau regions indicates that street $2$ can
maintain a constant flow capacity for a wide range of density
variations in the first street. The other feature is that in
fixed $\rho_1$, $J_2$ is an increasing function for small values
of $\rho_2$. This is natural since the current in street $2$ has
not reached its maximal value. This increment persists up to
$\rho_2=0.05$. Beyond that, for each $\rho_1$, $J_2$
saturates. In the plateau region, the saturation value is
slightly above $0.16$. The current saturation continues up to
$\rho_2=0.8$ above which $J_2$ again starts to decrease. We note
that the behaviours depicted in $J_1-\rho_1$ and $J_2-\rho_1$ diagrams
are consistent to each other. Due to the existence of $1
\rightleftharpoons 2$ symmetry, the $J_2-\rho_2$ diagram is
identical to $J_1-\rho_1$ and $J_1-\rho_2$ is identical to
$J_2-\rho_1$. In order to find a deeper insight,
it would be illustrative to look at the behaviour of total current
$J_{tot}=J_1+J_2$ as a function of density in one of the streets.
Fig. (5) sketches this behaviour.

\begin{figure}
\centering
\includegraphics[width=7cm]{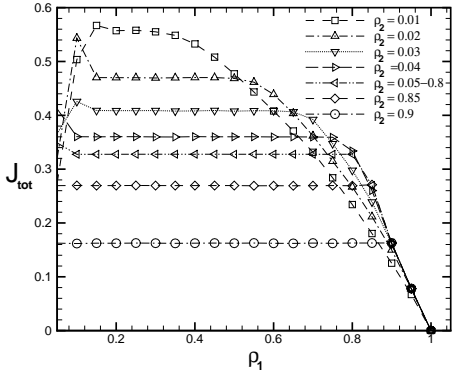}
\caption{ Total current $J_{tot}$ vs $\rho_1$ for various values of
$\rho_2$. $D_{s}=25~m$ and $L_{car}=5$ cells. } \label{fig:bz2}
\end{figure}

For $\rho_2<0.05$, the maximum of $J_{tot}$ lies at $\rho_1=0.1$.
However, for $\rho_2>0.05$, the maximum shifts backward to
$\rho_1=0$. According to the above graphs, after a short increasing behaviour, $J_{tot}$
enters into a lengthy plateau region. Evidently for optimisation of traffic
one should maximize the total current $J_{tot}$. The existence of
a wide plateau region in $J_{tot}$ suggests that yielding
mechanism can be regarded as an efficient method in the plateau
range of density in the first street. Let us now consider the role of $L_{car}$.
The cellular nature of our model permits us to adjust the cell length
$\Delta x$ in such a way to reproduce a reasonable acceleration.
Our simulations demonstrate that currents exhibit significant dependence
on $\Delta x$ or equivalently on $L_{car}$. This is exhibited in figures (6,7).

\begin{figure}
\centering
\includegraphics[width=7cm]{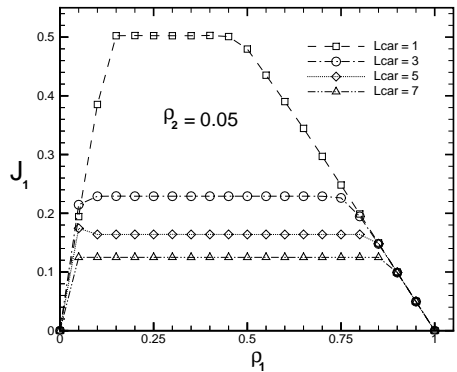}
\caption{ $J_1$ vs $\rho_1$ for various $L_{car}$. Global density of
the second street is kept fixed at $ \rho_2=0.05$. } \label{fig:bz2}
\end{figure}

While the structure of the fundamental diagram does not
qualitatively change, the values of $J_1$ notably depend on
$L_{car}$. For both $\rho_2=0.05$ and $0.5$, $J_1$ is a decreasing
function of $L_{car}$. The reason is that larger $L_{car}$ gives
rise to higher acceleration.

\begin{figure}
\centering
\includegraphics[width=7cm]{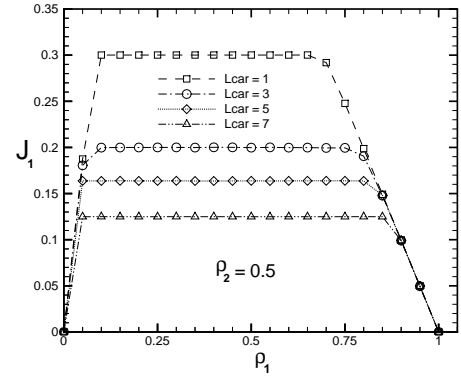}
\caption{ $J_1$ vs $\rho_1$ for various $L_{car}$ with $\rho_2=0.5$.
} \label{fig:bz2}
\end{figure}

Analogous to $J_1$, the dependence of $J_2$ on
$L_{car}$ is considerable as is shown in figures (8,9).
Variation of $L_{car}$ does not lead to change the generic behaviour but rather
changes the current values. Due to the same reason which was explained, smaller
$L_{car}$ gives higher currents. In the case $L_{car}=1$, the transition of $J_2$
from the plateau region to the linear decreasing segment is much smoother compared
to the other values of $L_{car}$ greater than one. Since the currents in the plateau
region do not depend on density, therefore the higher acceleration gives rise to
larger currents.

\begin{figure}
\centering
\includegraphics[width=7cm]{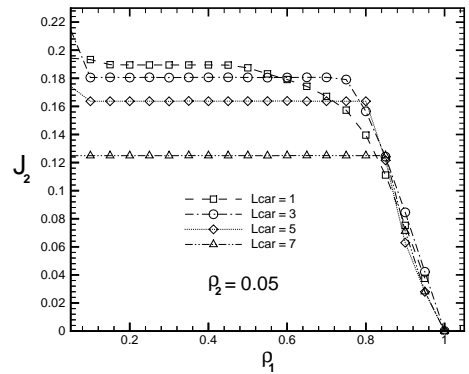}
\caption{ $J_2$ versus $\rho_1$ for various $L_{car}$.
$\rho_2=0.05$. } \label{fig:bz2}
\end{figure}

\begin{figure}
\centering
\includegraphics[width=7cm]{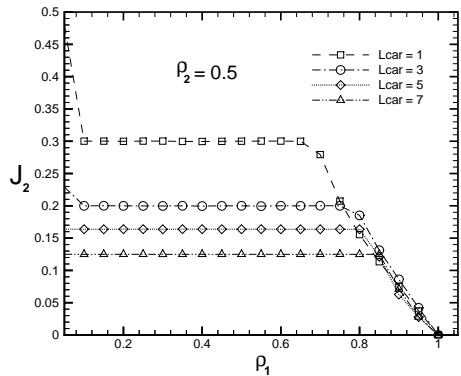}
\caption{ $J_2$ versus $\rho_1$ for various $L_{car}$. The density
is kept fixed in the second street at $\rho_2=0.5$ . }
\label{fig:bz2}
\end{figure}

Finally, we have also examined the effect of varying the safety
distance $D_s$. Our simulations do not show any significant dependence
on $D_s$. This is due to unrealistic decceleration in the NS model.

\section{Summary and Concluding Remarks}

We have investigated the flow characteristics of a
non signalised intersection by developing a Nagel-Schreckenberg
cellular automata model. In particular, we have obtained the
fundamental diagrams in both streets. Our findings show yielding of
cars upon reaching the intersection gives rise to
formation of plateau regions in the fundamental diagrams. This is
reminiscent of the conventional role of a single impurity in the
one dimensional out of equilibrium systems. The performance of
non-signalised controlling mechanism is especially efficient when
the car density is considerably low in both streets. The
existence of wide plateau region in the total system current
shows the robustness of the controlling scheme to the density
fluctuations and offers an optimal method for controlling the
traffic at low densities. Our CA model allows for varying space
and time grids. By their appropriate adjusting, we are
able to reproduce a realistic acceleration. In low densities, the
currents exhibits notable dependence on the values of spatial
discretisation grid. Finally we remark that our approach is open
to serious challenges. The crucial point is to model the yielding
braking as realistic as possible. Empirical data are certainly
required for this purpose. We expect the system characteristics
undergo substantial changes if realistic yielding declaration is
taken into account.

\section{acknowledgement}

We highly appreciate Kadkhodaa Yaghoub, Sardaar Kaamyaab and Mehdi
Neek Amal for their useful helps .

\end{document}